\documentclass[conference]{IEEEtran}

\begin{document}
\title{GPU-based Image Analysis on Mobile Devices}
\author{\IEEEauthorblockN{Andrew Ensor}
\IEEEauthorblockA{School of Computing and Mathematical Sciences\\
AUT University\\
Auckland, New Zealand\\
Email: andrew.ensor@aut.ac.nz}
\and
\IEEEauthorblockN{Seth Hall}
\IEEEauthorblockA{School of Computing and Mathematical Sciences\\
AUT University\\
Auckland, New Zealand\\
Email: sehall@aut.ac.nz}}

\maketitle

\begin{abstract}
With the rapid advances in mobile technology many mobile devices are capable of
capturing high quality images and video with their embedded camera.
This paper investigates techniques for real-time processing of the resulting images,
particularly on-device utilizing a graphical processing unit.
Issues and limitations of image processing on mobile devices are discussed,
and the performance of graphical processing units on a range of devices measured
through a programmable shader implementation of Canny edge detection.
\end{abstract}

\section{Introduction}

Mobile phone technology is virtually ubiquitous and rapidly evolving, giving rise to
new and exciting application domains through the convergence of communication,
camera and computing technologies.
Many of these applications, such as those for mobile augmented reality,
utilize the device camera for image recognition or visual tag identification
\cite{Bib:SantosSierra}, \cite{Bib:Karodia}, \cite{Bib:Lee},
\cite{Bib:ARToolKit}.
Mobile devices have quite distinct capabilities and limitations from
desktop computers,
so many of the usual approaches for application development
must be reworked to be made suitable for deployment to actual mobile devices.
For instance, the procedure for capturing images varies from device to device,
and the quality, contrast, resolution and rates of image capture can be
substantially different.
The central processing unit capabilities of many devices is a significant inhibiting
factor for realizing some applications, as can be the network communication bandwidth, latency, and cost, as well as demands on the finite battery charge.

However, mobile computational capabilities and memory specifications are rapidly
evolving making more processor-intensive applications possible that were considered
infeasible even two years ago.
For instance, the Nokia N series of multimedia devices commenced with the release
of the Nokia N70 in 2005, which included a $2$ megapixel rear camera
(and $0.3$ megapixel front camera), $32$ MB memory, and a $220$ MHz ARM-926 CPU.
In 2007 the Nokia N95 was released with a $5$ megapixel rear camera, $160$ MB memory,
and a $330$ MHz ARM-11 CPU.
More recently, the Nokia N8 was released in 2010 with a $12$ megapixel rear camera,
$256$ MB memory, and both a $680$ MHz ARM-11 CPU and a BCM2727 GPU capable of $32$ MPoly/s.
It is now common for newer smart phones to include a $1$ GHz CPU and a GPU such as a
PowerVR SGX (Imagination Technologies), Adreno (Qualcomm, formerly of AMD), Mali (ARM),
or Tegra 2 (NVIDIA).

\section{Image Capture and Analysis on Mobile Devices}

Images can be obtained by an application from a mobile camera by
taking a photograph snapshot. However, this can be a notoriously slow process,
requiring between $520$ ms and $8$ s for some N-series devices \cite{Bib:Gu}.
Instead, it is far preferable to obtain preview frames from the video.
On Java ME supported mobiles the commonly available Multimedia API provides access
to video data. However, device implementations of this API usually require that
the video capture be stopped to obtain and then separately decode the video segment
(typically in 3GPP format) in order to obtain any frames.
Some platforms, such as Android, allow both RGB and greyscale preview frames to be captured
(with typical rates for a $640\times480$ image of $26$ frames per second
on a Google Nexus One and $30$ frames per second on an HTC Desire HD),
whereas others, such as iOS, only return RGB frames by default
(with typical rates of 29 frames per second on an Apple iPhone 4)
which can then be converted by software to greyscale if necessary for further analysis.

Once captured there are two (non-exclusive) choices for processing an image:
\begin{itemize}
  \item \emph{off-device}
  utilizing the network capabilities of the mobile,
  either a localized network technology such as Bluetooth or Wi-Fi,
  or a cellular network to off-load the image processing to a more
  powerful machine,

  \item \emph{on-device}
  utilizing the computing capabilities of the mobile to itself
  perform the processing via the CPU or GPU.
\end{itemize}
For instance, the \emph{Shoot \& Copy} application \cite{Bib:Boring1}
utilizes Bluetooth to pass a captured image to a Bluetooth server for identification
and contextual information about the image.
The \emph{Touch Projector} application \cite{Bib:Boring2}
passes video and touch events via Wi-Fi to a computer connected to a projector.
However, off-device processing has some significant disadvantages.
Although many devices support Bluetooth 2.0 with enhanced data rates providing a
theoretical data transfer rate of $2.1$ Mbps, the authors found that in practice on most
devices the rate was closer to $430$ kbps upload and $950$ kbps download,
which can result in a significant communication latency when transmitting image frames.
Wi-Fi improves the bandwidth and reduces latency but it has somewhat less support
on older mobile devices and can be quite demanding on the battery.
Whereas both Bluetooth and Wi-Fi are only suitable for localized processing
solutions, utilizing a cellular network with a persistent but mostly idle
TCP connection to a processing server can provide a more suitable off-device
solution.
However, this too can result in significant network-specific bandwidth limitations
(a 3G network has typical speeds of $150$ kbps upload and $2$ Mbps download),
latencies, and usage charges.
The eventual availability of LTE promises to reduce this issue with $50$ Mbps upload,
$100$ Mbps download, and round trip latencies reduced to around $10$ ms.

With the evolving specifications of mobile devices there is a growing list of
literature and applications that choose to perform image processing on-device.
On-device processing was used by \cite{Bib:Reitmayr} for edge-based tracking
of the camera pose by a tablet PC in an outdoor environment.
PhoneGuide \cite{Bib:Bruns} performed object recognition computations on a mobile phone.
SURF \cite{Bib:Bay} was implemented on a Nokia N95 to match camera images against
a database of location-tagged images
\cite{Bib:Takacs} providing image matches in 2.8 seconds.
Variants of SIFT and Ferns algorithms were used in
\cite{Bib:Wagner1}, and \cite{Bib:Wagner2}
tested them on an Asus P552W with a 624 MHz Marvell PXA 930 CPU
with the algorithms processing a $240\times320$ frame in $40$ ms.
Studierstube ES \cite{Bib:Schmalstieg} is a marker tracking API that is a successor to
ARToolKitPlus and available for Windows CE, Symbian, and iOS, but it is closed source.
Junaio 3.0 \cite{Bib:Junaio} is a free augmented reality browser for
iOS and Android platforms that utilizes image tracking to display objects
from a location-based channel (showing points of interest in surroundings)
or a Junaio GLUE channel (attaching virtual 3D models to up to seven visible markers).
Most other mobile applications, such as Google Goggles \cite{Bib:Googles} for Android and iOS
have entirely web-based pattern matching so no image analysis is performed on the device.
From version 2.2 the popular OpenCV API \cite{Bib:OpenCV} has been available for
Android and Maemo/Meego platforms, and it also can be built for iOS.
NVidia has contributed (non-mobile) GPU implementations of some computer vision algorithms,
and has contributed optimizations for the Android CPU implementation.

It is now commonplace for applications to utilize GPU for processing beyond
only graphics rendering,
particularly for tasks that are highly parallel and have high arithmetic intensity,
for which GPU are well suited.
As most computer vision algorithms take an array of pixel data as input and output
a variable-length representation of the image (the reverse of graphics rendering
for which GPU were originally designed) their implementation on GPU has been somewhat slower
than by some other fields.
Some examples of computer vision algorithms implemented on GPU can be found in
\cite{Bib:Fung}, \cite{Bib:Allusse}, and \cite{Bib:Junchul}.
However, mobile devices containing programmable GPU only became widely available in 2009
with the use of the PowerVR SGX535 processor, so to date there has been very little
literature available on mobile-specific GPU implemented algorithms.
Several recent articles and potential power savings by utilizing GPU rather than CPU on mobiles
are discussed in \cite{Bib:Kwang}.
In particular, \cite{Bib:Singhal} implements a Harris corner detection on a
OMAP ZOOM Mobile Development Kit equipped with a PowerVR SGX 530
GPU using four render passes
(greyscale conversion, gradient calculations, Gaussian filtering and corner strength calculation,
and local maxima), reporting $6.5$fps for a $640\times480$ video image.

\section{OpenGL ES}

With the notable exception of Windows Phone devices the vast majority of modern mobile
devices support OpenGL ES, a version of the OpenGL API that is intended for embedded systems.
From version 2.0 OpenGL ES supports programmable shaders, so parts of an application can
be written in GLSL and executed directly in the GPU pipeline.

As with all shaders branching is discouraged as it carries a performance penalty,
particularly when it involves dynamic flow control on a condition computed within each
shader, although the shader compiler may be able to compile out static flow
control and unroll loops computed on compile-time constant conditions or uniform variables.
The reason for this is that GPU don't have the branch-prediction circuitry that is
common in CPU,
and many GPU execute shader instances in parallel in lock-step,
so one instance caught inside a condition with a substantial amount of computation
can delay all the other instances from progressing.
The same holds for dependent texture reads, where the shader itself computes texture coordinates
rather than directly using unmodified texture coordinates passed into the shader.
The graphics hardware cannot then prefetch texel data before the shader executes
to reduce memory access latency.
Unfortunately, many computer vision algorithms require dependent texture reads when
implemented on a GPU.
Another issue that must be considered is the latency in creating and transferring textures.
Ideally, all texture data for a GPU should be loaded during initialization and
preferably not changed while the shaders execute,
to reduce the dataflow between memory and the GPU.
However, for real-time image analysis to be feasible on a GPU image data captured from
the camera should preferably be loaded into a preallocated texture at 30 fps,
quite contrary to GPU recommended practices.
This can be partially compensated for by reducing the image resolution or
changing its format from RGB vector float values to integer or compressed.

OpenGL ES 2.0 allows byte, unsigned byte, short, unsigned short, float, and fixed
data types for vertex shader attributes, but vertex shaders always expect attributes
to be float so all other types are converted, resulting in a compromise between
bandwidth/storage and conversion costs.
It requires that a GPU must allow at least two texture units to be available to
fragment shaders, which is not an issue for many image processing algorithms,
although most GPU support eight texture units.
Textures might not be available to vertex shaders and there are often tight limits
on the number of vertex attributes and varying variables that can be used
($16$ and $8$ respectively in the case of the PowerVR SGX series of GPU).

Unlike the full version OpenGL ES uses precision hints for all shader values:
\begin{itemize}
  \item lowp for $10$ bit values between $-2$ and $1.999$ with a precision of $1/256$
  (which for graphics rendering is mainly used for colours and reading from low precision
  textures such as normals from a normal map),

  \item mediump for $16$ bit values between -65520 and 65520 consisting of a sign bit,
  $5$ exponent bits, and $10$ mantissa bits
  (which can be useful for reducing storage requirements),

  \item highp for 32 bit (mostly adhering to the IEEE754 standard).
\end{itemize}
Furthermore, the GPU on a mobile device is most likely to be a scalar rather than
vector processor. This means that there is typically no advantage vectorizing highp
operations, as each highp component will be computed sequentially,
although lowp and mediump values can be processed in parallel.
It is also common for GPU on mobiles to use tile-based deferred rendering,
where the framebuffer is divided into tiles and commands get buffered and processed
together as a single operation for each tile.
This helps the GPU to more effectively cache framebuffer values and allows it
to discard some fragments before they get processed by a fragment shader
(for this to work correctly fragment shaders should themselves avoid discarding fragments).

There are performance benchmarks for the GPU commonly found in mobile devices
\cite{Bib:GlBenchmark}.
However, the benchmarks typically only compare the performance for graphics rendering throughput,
not for other tasks such as image processing, so do not significantly test the
implications of effects such as frequent texture reloading and dependent texture reads.

\section{Canny Shader Implementation}\label{Sec:Canny}

Canny edge detection \cite{Bib:Canny} is one of the most commonly used
image processing algorithms,
and it illustrates many of the issues associated with implementing image processing
algorithms on GPU.
It has a texture transfer for each frame captured,
a large amount of conditionally executed code, and dependent texture reads.
As such it might not be considered an ideal candidate for implementation on a GPU.

The Canny edge detection algorithm is based on the gradient vector and can give
excellent edge detection results in practice.
Starting with a single channel (greyscale) image it proceeds in four steps to produce
an image whose pixels with non-zero intensity represent the edges in the original image:
\begin{itemize}
  \item First the image is smoothed using a Gaussian filter to reduce some of the noise.

  \item At each pixel in the smoothed image the gradient vector is calculated using the
  two Sobel operators. The length $\left|\nabla f\right|$ of the gradient
  vector is calculated or approximated, and its direction is classified into one of the
  four directions horizontal, vertical, forward diagonal, or backward diagonal
  (depending to which direction $\nabla f$ is closest).

  \item At each pixel \emph{non-maximum suppression} is applied to the value of
  $\left|\nabla f\right|$ by comparing the value of $\left|\nabla f\right|$ at the pixel
  with its value at each of the two opposite neighbouring pixels in either direction.
  If its value is smaller than the value at either of those two pixels then the pixel
  is discarded as not a potential edge pixel (value is set to $0$ as the neighbouring
  pixel has a greater change in intensity so it better represents the edge).
  This results in \emph{thin lines} for the edges.

  \item At each remaining pixel a \emph{double threshold}
  (or \emph{hysteresis threshold}) is applied using both an upper and a lower threshold,
  with a ratio upper:lower typically between 2:1 and 3:1.
  If the pixel has a value of $\left|\nabla f\right|$ above the upper threshold
  then it is accepted as an edge pixel (and referred to as a \emph{strong pixel}),
  whereas any pixel for which $\left|\nabla f\right|$ is below the lower threshold is rejected.
  For any pixel whose value of $\left|\nabla f\right|$ is between the upper and lower thresholds,
  it is accepted as an edge pixel if and only if one of its eight neighbours is above the threshold
  (it has a strong pixel neighbour, in which case the pixel is referred to as a \emph{weak pixel}).
\end{itemize}

Canny edge detection was implemented in \cite{Bib:Ogawa}
using CUDA on a Tesla C1060 GPU with $240$ $1.3$ GHz cores.
The GPU implementation achieved a speedup factor of 50 times over a conventional
implementation on a $2$ GHz Intel Xeon E5520 CPU, although both these GPU and CPU were
far more powerful than the processors currently found in mobile devices.

In this work the authors have created a purely GPU-based implementation of the
Canny edge detection algorithm and tested its performance across a range of popular
mobile devices that support OpenGL ES 2.0 using the camera on each device.
The purpose is to determine whether it is yet advantageous to utilize the GPU
in these devices for image analysis instead of the usual approach of having the
processing performed entirely by the CPU.
To achieve this the algorithm was implemented in GLSL via a total of five render passes
using four distinct fragment shaders all having mediump precision:
\begin{itemize}
  \item Gaussian smoothing using either a $3\times3$ or a $5\times5$ convolution kernel.
  Since a Gaussian kernel is separable it can be applied as two one-dimensional
  convolutions so the Gaussian smoothing is performed in two passes,
  trading the overhead of a second render pass against the lower number of texture reads.
  Even for a $3\times3$ kernel using two render passes rather than one was found to benefit
  performance on actual devices.

  \item The gradient vector is calculated and its direction is classified.
  First the nine smoothed pixel intensities are obtained in the neighbourhood of a pixel,
  and used by the Sobel X and Y operators to obtain the gradient vector.
  Then IF statements are avoided by multiplying the gradient vector by a $2\times2$
  $\frac{1}{16}$-turn rotation matrix and then its angle relative to horizontal is doubled
  so that it falls into one of four quadrants.
  A combination of step and sign functions is then used to classify the
  resulting vector as one of the eight primary directions $(\Delta_x,\Delta_y)$
  with $\Delta_x$ and $\Delta_y$ each being either $-1$, $0$, or $1$.
  These eight directions correspond to the four directions in the usual
  Canny edge detection algorithm along with their opposite directions.
  The shader then outputs the length of the gradient vector and the vector
  $(\Delta_x,\Delta_y)$.
  This approach to classifying the direction was found to take as little as half the
  time of several alternative approaches that utilized conditional statements.

  \item Non-maximal suppression and the double threshold are applied together.
  Non-maximal suppression is achieved by obtaining the length of the gradient vector
  from the previous pass for the pixel with the length of the gradient vector
  for the two neighbouring pixels in directions $(\Delta_x,\Delta_y)$
  and $(-\Delta_x,-\Delta_y)$.
  The length at the pixel is simply multiplied by a step function that
  returns either $0.0$ or $1.0$ depending whether its length is greater
  than the maximum of the two neighbouring lengths.
  For the double threshold a smoothstep is used with the two thresholds
  to output an edge strength measurement for the pixel between $0.0$ (reject)
  and $1.0$ (accept as a strong pixel).

  \item The final shader handles the weak pixels differently from
  Canny's original algorithm.
  Rather than simply accepting a pixel as a weak pixel if one of its neighbouring
  eight pixels is a strong pixel, since the previous render pass has provided
  an edge strength measurement for each pixel more information is available.
  This shader obtains the nine edge strength measurements in the neighbourhood
  of a pixel, and takes a linear combination of the edge strength measurement at
  the pixel with a step function that accepts a weak pixel if the sum of the
  nine edge strength measurements is at least $2.0$.
  This avoids the usual IF statement with eight OR conditions,
  greatly increasing performance of this render pass and giving a small improvement
  in the weak pixel criterion.
\end{itemize}
In effect, the entire Canny edge detection algorithm is implemented without
any conditional statements whatsoever, ideal for a GPU shader-based implementation
on OpenGL ES.
The shader code is available from the authors upon request.

\section{Performance Results}\label{Sec:PerformanceResults}

The GPU version of the Canny edge detection described in Section \ref{Sec:Canny}
was implemented on the following devices, chosen as they were all released within the
same year and now commonplace:
\begin{itemize}
  \item Google Nexus One,
  released January 2010,
  operating system Android 2.3,
  CPU 1 GHz Qualcomm QSD8250 Snapdragon,
  GPU Adreno 200,
  memory 512 MB RAM,
  camera 5 megapixel,
  video $720\times480$ at minimum $20$ fps.

  \item Apple iPhone 4,
  released June 2010,
  operating system iOS 4.3.5,
  CPU Apple A4 ARM Cortex A8,
  GPU PowerVR SGX 535,
  memory 512 MB RAM,
  camera 5 megapixel,
  video 720p ($1280\times720$) at $30$ fps.

  \item Samsung Galaxy S,
  released June 2010,
  operating system Android 2.3,
  CPU 1 GHz Samsung Hummingbird S5PC110 ARM Cortex A8,
  GPU PowerVR SGX 540 with 128 MB GPU cache,
  memory 512 MB RAM,
  camera 5 megapixel,
  video 720p at $30$ fps.

  \item Nokia N8,
  released September 2010,
  operating system Symbian\verb|^|3,
  CPU 680 MHz Samsung K5W4G2GACA-AL54 ARM 11,
  GPU Broadcom BCM2727,
  memory 256 MB RAM,
  camera 12 megapixel,
  video 720p at $25$ fps.

  \item HTC Desire HD,
  released October 2010,
  operating system Android 2.3,
  CPU 1 GHz Qualcomm MSM8255 Snapdragon,
  GPU Adreno 205,
  memory 768 MB RAM,
  camera 8 megapixel,
  video 720p at $30$ fps.

  \item Google Nexus S,
  released December 2010,
  operating system Android 2.3,
  CPU 1 GHz Samsung Hummingbird S5PC110 ARM Cortex A8,
  GPU PowerVR SGX 540,
  memory 512 MB RAM,
  camera 5 megapixel,
  video $800\times480$ at $30$ fps (not 720p).
\end{itemize}
The Android devices directly supported obtaining the video preview
in YUV format, and the Y component could be used as input as a greyscale image
without the requirement for any preliminary processing.
However, the iOS and Symbian\verb|^|3 devices only supported obtaining the preview in RGB,
so they required an additional preliminary render pass to convert the
RGB image to greyscale.
An additional point worth mentioning for the iPhone is that any pending
OpenGL ES commands must be flushed before the application is put into the background,
otherwise the application gets terminated by the operating system.

Table \ref{Tab:RenderPassTimes} lists the times in milliseconds for each of the
render passes for some of the devices.
\begin{table}[!t]
\renewcommand{\arraystretch}{1.3}
\caption{Render Pass and Image Reloading Texture Times (ms)}
\label{Tab:RenderPassTimes}
\centering
\begin{tabular}{|l|rrr|}\hline
Operation   & Nexus One     & iPhone 4     & Desire HD    \\\hline
Greyscale   & n/a           & $8.9\pm3.0$  & n/a          \\
Gaussian X  & $29.9\pm4.9$  & $12.2\pm0.8$ & $11.1\pm3.3$ \\
Gaussian Y  & $29.0\pm4.5$  & $12.0\pm0.1$ & $11.2\pm3.7$ \\
Gradient    & $138.2\pm3.9$ & $60.2\pm0.4$ & $22.5\pm1.4$ \\
Non-max Sup & $50.1\pm6.0$  & $25.1\pm2.7$ & $11.2\pm1.8$ \\
Weak Pixels & $78.8\pm2.5$  & $28.9\pm4.4$ & $19.7\pm1.0$ \\
Reload texture & $86.6\pm12.8$ & $36.8\pm4.3$ & $5.2\pm4.8$  \\\hline
\end{tabular}
\end{table}
To obtain these times the OpenGL ES glFinish command was used to flush
any queued rendering commands and wait until they have finished.
Note this removes the ability of the GPU to commence further commands,
so although useful for comparing the times required for each render pass,
their sum only gives an upper bound on the total algorithm time.
The two Gaussian smoothing render passes were timed using a $3\times3$ convolution kernel.
Using instead a Gaussian $5\times5$ kernel was found to add between an extra
$3$ ms (for iPhone 4 and Desire HD) and an extra $10$ ms (Nexus One)
to each of the two Gaussian render passes,
but did not have any visibly noticeable effect on the edge detection results.
The calculation of the gradient vector is the most burdensome render pass,
explained by the nine texture reads it performs and relatively
complex computation used to classify its direction. This number of texture reads
is also performed in the weak pixels render pass, whereas the other two render passes
only require three texture reads.
The table also gives the time required to copy captured image data to the texture,
which is an important quantity for real-time processing of images captured from
the device camera, and dictated by the GPU memory bandwidth.
A $640\times480$ (VGA, non-power-of-two) image was used,
a common resolution available for video preview on all the devices,
although most supported greater resolutions as well.
No texture compression was used which would introduce conversion
latency but assist texture data to better fit on the memory bus and in a texture cache.

The results in Table \ref{Tab:FrameRates} show the actual overall frame rates that
were achieved in practice on each device.
As the OpenGL ES glTexImage2D command used to update a texture with new image data
blocks until all the texture data has been transferred,
for efficiency the (non-blocking) render pass commands were performed before glTexImage2D
was called to set the texture with a image capture for the next set of render passes
--- this was found to help increase frame rates.
\begin{table}[!t]
\renewcommand{\arraystretch}{1.3}
\caption{Frame Rates for Image Capture and Edge Detection (fps)}
\label{Tab:FrameRates}
\centering
\begin{tabular}{|l|rrr|}\hline
Device    & CPU+Android Cam & CPU+Native Cam & GPU Shaders \\\hline
Nexus One & $7.5\pm1.8$ & $9.7\pm0.7$  & $3.9\pm0.2$ \\
iPhone 4  & n/a         & $7.4\pm0.4$  & $7.6\pm0.0$ \\
Galaxy S  & $9.1\pm0.5$ & $14.8\pm0.1$ & $11.3\pm0.2$ \\
Nokia N8  & n/a         & n/a          & $14.5\pm0.1$ \\
Desire HD & $7.1\pm1.3$ & $10.7\pm0.8$ & $15.4\pm0.2$ \\
Nexus S   & $8.2\pm0.9$ & $15.5\pm0.8$ & $8.9\pm0.4$ \\\hline
\end{tabular}
\end{table}
To provide some comparison with the CPU performance on each device, an OpenCV version
of Canny edge detection was also timed (unlike the iOS build of OpenCV, the Android version
currently has an optimized platform-specific build available).
No specific Symbian\verb|^|3 release of OpenCV was available during testing.
As the OpenCV edge detection relies on the performance of the CPU,
wherever practical any applications running in the background on the device were stopped.
On the Android devices it was found that the burden on the CPU associated with
obtaining an image capture could be significantly reduced by using a native camera
capture API rather than the default Android API, hence the two sets of CPU results
reported.

\section{Discussion and Conclusions}

Perhaps the most interesting conclusion that can be drawn from the results in
Section \ref{Sec:PerformanceResults} is the great variation in the ability
of different GPU in the mobile market for performing image processing.
The Nexus One with an Adreno 200 GPU displayed quite poor performance,
due to the time to transfer texture data and its slower execution of shader code.
However, the Desire HD with the newer Adreno 205 GPU provided surprisingly
good results, receiving at least a $50\%$ performance benefit by offloading
edge detection to the GPU rather than CPU. Both these devices use Snapdragon CPU
which were seen to execute OpenCV code slower than their competing Hummingbird CPU,
found on the Galaxy S and Nexus S. For these two devices the benefit of running the
edge detection on the GPU is less definitive, although doing so would free up the
CPU for other processor-intensive tasks that might be required by an application.
The GPU results for the N8 with its Broadcom GPU were encouraging as its processor
hardware is common across Symbian\verb|^|3 devices of the era,
whereas the GPU results for the iPhone 4 are not surprising,
it uses an older PowerVR SGX535
rather than the newer PowerVR SGX540 found in the Galaxy S and Nexus S.
It should be reiterated that the iPhone CPU results were taken using an OpenCV
build that was not optimized for that platform.

It is worthwhile to compare the frame rates with some of the OpenGL ES rendering
benchmarks that are available.
For instance, \cite{Bib:GlBenchmark} reports comparative benchmark results for
Nexus One ($819$), iPhone 4 ($1361$), Galaxy S ($2561$), Desire HD ($2377$),
and Nexus S ($2880$).
These results do depart somewhat from the GPU fps results in
Section \ref{Sec:PerformanceResults},
indicating differences between benchmarking GPU for typical graphics rendering
versus performing an image processing algorithm such as Canny edge detection.

The general pattern in the GPU ability for image processing appears to have reached
a tipping point during the 2010 release period of the investigated devices,
with some devices clearly being able to benefit from offloading processing to the GPU.
As GPU continue to rapidly evolve, with the release of Adreno 220 and PowerVR SGX543,
along with new GPU such as the Mali and the Tegra 2 for mobile
devices available on devices in 2011, this benefit is only continuing to increase.
For instance, modest performance improvements are observed in the Sony Ericsson Xperia Arc,
released in April 2011 with same CPU and GPU as the Desire HD,
with the CPU+Android Camera tests achieving $10.0\pm1$fps
and GPU shaders achieving $17.5\pm0.1$fps.
More impressive are the results for the Samsung Galaxy S2, first released in May 2011
with a 1.5 GHz Snapdragon S3 CPU and Mali-400 GPU.
Its CPU+Android Camera tests achieved $14.2\pm0.7$fps,
which were dwarfed by the GPU shader results of $33.8\pm3.6$fps.


%\bibliographystyle{IEEEtran}

\end{document}